\documentclass[aps, twocolumn]{revtex4-1}

\usepackage{amsmath,amssymb,mathrsfs}

\usepackage{graphicx}
\usepackage{color}
\usepackage{hyperref}

\def\be{\begin{equation}}
\def\ee{\end{equation}}
\def\bea{\begin{eqnarray}}
\def\eea{\end{eqnarray}}
\def\nn{\nonumber}

\def\tabOhclass{$1$}

\def\npsection#1{{\par{\vskip 5pt}\noindent\bf #1}}

\def\papertitle{Exotic topological density waves in cold atomic Rydberg fermions}

\begin{document}

\title{\papertitle}
\author{Xiaopeng Li, S. Das Sarma} 
\affiliation{Condensed Matter Theory Center and Joint Quantum Institute, Department of Physics, University of Maryland, College Park, MD 20742-4111, USA}

\maketitle

{\bf 
Versatile controllability of interactions in ultracold atomic and molecular gases has now reached an unprecedented era where quantum correlations and unconventional many-body phases can be studied with no corresponding analogs in solid state systems. Recent experiments in Rydberg atomic gases have achieved exquisite control over non-local interactions~\cite{2009_Low_Rydberg_PRA,2012_Bloch_Rydberg_Nature,2014_Bloch_Rydberg_arXiv,2014_Low_Rydberg_News,2014_Browaeys_NatPhys}, allowing novel quantum phases unreachable with the usual local interactions in atomic systems. Here, we study Rydberg dressed atomic fermions in a three dimensional optical lattice predicting the existence of hitherto unheard-of exotic mixed topological density wave phases.  
We show that varying spatial range of the non-local interaction leads to a rich phase diagram containing various bond density waves,  with unexpected spontaneous time-reversal symmetry breaking. Quasiparticles in these chiral phases  experience emergent gauge fields and form three dimensional quantum Hall and Weyl semimetal states. Remarkably, certain density waves even exhibit mixed topologies beyond the existing topological classification. 
Experimental signatures of density waves and their topological properties are predicted in time-of-flight measurements. 
} 

Topological quantum states of matter have attracted considerable theoretical and experimental attention in condensed matter research in the last decade, aiming at robust physical properties protected (often due to some underlying symmetry) against variations in microscopic details. For non-interacting fermions, topology of gapped phases has been classified for arbitrary dimensions~\cite{2009_Kitaev_arXiv,2010_Ryu_NJP}. Examples are Z class  Chern insulators in two dimensions manifesting quantum anomalous Hall effect~\cite{1988_Haldane_QHE,2013_Xue_QAHE}, and time-reversal invariant Z$_2$ topological insulators in three dimensions with robust magnetoelectric response~\cite{2007_Fu_PRL,2007_Moore_PRB,2008_Qi_Zhang_PRB,2009_Essin_Moore_Vanderbilt_TI}. For classification of gapless systems, no unified theoretical framework has emerged yet, nonetheless recent studies on semimetals suggest one theoretical route by characterizing topological properties of band touching points, such as flux monopoles~\cite{2011_Leon_PRL,2011_Ashvin_Weyl_PRB,2011_Dai_Weyl_PRL,2012_Sun_NatPhys,2014_Yang_NatComm}. In this paper, we provide one fascinating possibility beyond the present theoretical paradigm in an interacting system, with mixed gapped and gapless topologies emergent in a three dimensional bond density wave state. We demonstrate the existence of such an exotic topological  density wave in a Rydberg dressed atomic Fermi gas via microscopic calculations. We also propose an experimental scheme to extract the topological properties based upon time-of-flight signals.

We consider Rydberg dressed atomic Fermi gas in a three dimensional (3d) cubic optical lattice. This Rydberg atomic system has density-density interactions described by a non-local potential, $
V({\bf r}) = {V_6} /[{1+(|{\bf r}|/r_c)^6}], 
$ 
where $V_6$ describes the interaction strength, and $r_c$  the interaction range determined by the Condon  radius  in Rydberg dressing~\cite{2010_Henkel_PRL,2014_Zoller_Rydberg_arXiv}. Considering a deep lattice, the kinetic motion of atoms arises mainly from quantum tunneling between nearest neighboring sites providing  a unique kinetic energy scale, $t$. We will focus on a lattice with octahedral $O_h$ symmetry. The intrinsic properties of this system will depend on only two dimensionless parameters---$V_6/t$ and $r_c/a$ (with $a$ the lattice spacing), representing the strength of interaction and its non-locality, respectively.

At half filling, namely one atom per two lattice sites, the Fermi surface is perfectly nested in the deep lattice limit, because the single particle energy dispersion satisfies 
$ 
\epsilon _{ {\bf k} + {\bf Q} }= -\epsilon_{\bf k}, 
$
with ${\bf Q} = (\pi, \pi, \pi)$ (see Methods). 
The susceptibility towards forming particle-hole pairing or density waves 
$$
\rho_{\bf k} = \langle \psi ^\dag ({\bf k} + {\bf Q})   \psi ({\bf k} ) \rangle, 
$$
(with $\psi({\bf k})$ a Fourier transformed annihilation operator)  
is logarithmically divergent due to nesting and the system has generic  instabilities with repulsive interactions. With nearest neighbor interactions for spinless fermions, the ground state  
density wave order $\rho_{ \bf k }$ is momentum independent, implying a trivial 3d checkerboard pattern in real space.  
However, for Rydberg dressed atoms, the interaction range, $r_c$ can be comparable to several lattice spacings, which then frustrates the checkerboard pattern, giving rise to possibilities of unconventional density waves~\cite{2000_Nayak_PRB,2012_Zhao_PRL}.

\begin{figure*}[htp] 
\includegraphics[angle=0,width=\linewidth]{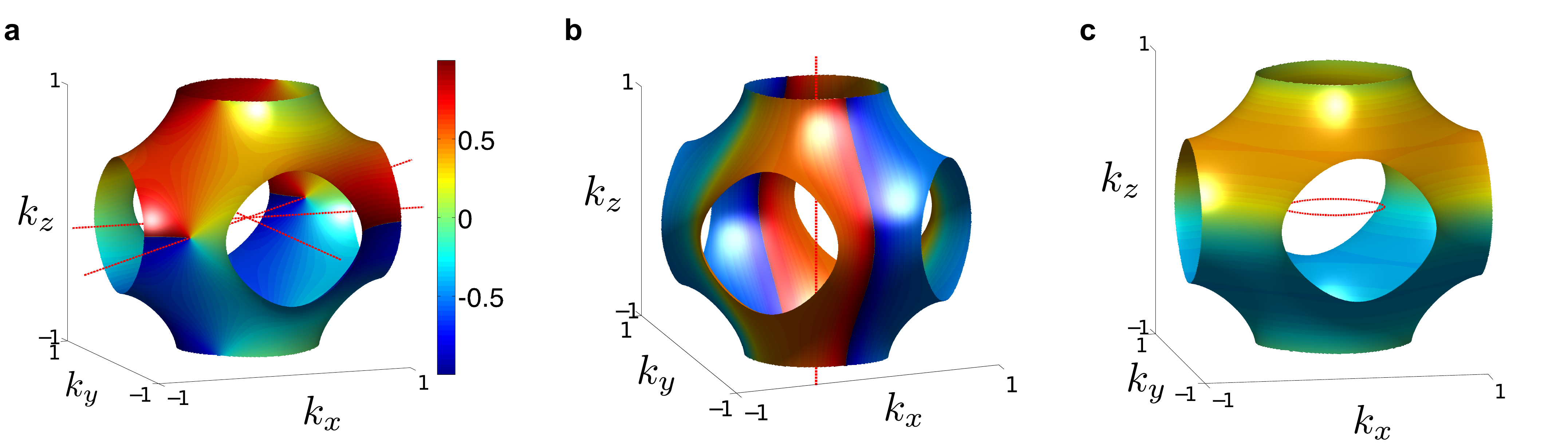}
 \caption{Density wave topology. The color index the phase angle of the momentum dependent density wave order $\Delta_{\bf k}$ projected on the Fermi surface. {\bf b} and {\bf c} use the the same color scheme as shown in {\bf a}.  {\bf a},  {\bf b} and {\bf c} correspond to topologically distinct density wave states. 
 The red dotted lines are  vortex lines around which the phase of $\Delta_{\bf k}$ changes by $2m\pi$ ($m\neq 0$). Right on vortex lines, $\Delta_{\bf k}$ vanishes. In {\bf a},  one vortex line crosses the Fermi surface. In both {\bf a} and {\bf b}, all vortex lines cannot be adiabatically removed without touching the Fermi surface, and thus are topologically robust, whereas in {\bf c}, the vortex line is adiabatically removable and is thus topologically trivial. 
}
\label{fig:topology} 
\end{figure*}

Within one-loop renormalization group (RG) analysis, we find that the strength of density wave instability is determined by the eigenvalue problem of a rescaled interaction matrix $\gamma$ whose symmetry is $O_h \times {\cal T}$ (see Methods).  Different density waves correspond to eigenvectors of $\gamma$, which are classified according to irreducible representations of the symmetry group~\cite{GroupTheory}. The symmetry classification of density waves and their representative irreducible basis functions are shown in Supplementary Table~\tabOhclass. 
The trivial density wave is $A_{1g}^+$ and all other density waves necessarily have more complicated momentum dependence for symmetry reasons. The latter leads to particle-hole pairing on the bonds in real space~\cite{2000_Nayak_PRB}.

Before presenting our microscopic results,  let us first look at  a nontrivial density wave, ${T_{1u} ^-} + T_{2u}^+ $,  which has a complex order $\Delta_{\bf k}$ (see Methods) of the form, 
\bea 
\Delta_{\bf k}& =&   
	\Delta_{T_{2u}^+} \left[ \sin k_x ( \cos k_y - \cos k_z ) + \sin k_y (\cos k_z - \cos k_x)   \right] \nn \\ 
	&& 		+ i \sqrt{2} \Delta_{T_{1u} ^-} \sin k_z . \nn
\eea 
Here $\Delta_{T_{1u} ^-}$ and $\Delta _{T_{2u}^+}$ take real values. Such a density wave order has a  rich topological structure in 3d  momentum space (Fig.~\ref{fig:topology}a), with three vortex lines located at $( 0, l, 0)$, $(l, 0, 0)$ and $(l, l, 0)$, with $l \in [-\pi/a,  \pi/a)$.  These vortex lines are topologically robust in the sense that they cannot be smoothly removed without touching the Fermi surface. The Bogoliubov-de Gennes Hamiltonian to describe  quasiparticles in the  density wave background reads,  
\be 
H_{\rm BdG} ({\bf k}) 
= 
\left[ 
\begin{array}{cc} 
\epsilon_{\bf k}	& \Delta_{\bf k} \\ 
\Delta ^*_{\bf k} 	& 	-\epsilon_{\bf k} 
\end{array} 
\right],  
\label{eq:BdGHam} 
\ee 
which can be rewritten in terms of Pauli matrices as  
$ 
H_{\rm BdG}  ({\bf k}) = \vec{ h } ({\bf k}) \cdot \vec{\sigma}. 
$  
At the points ${\bf k}_0^{\pm } =  ( \pm 2\pi/3,\pm 2\pi/3,  0)$ where vortex lines cross the Fermi surface, 
the magnitude of $\vec{h}$ vanishes, leading to zero energy quasiparticles. 
Around each nodal point, say  
${\bf k} = {\bf k}_0 ^+ + ( \delta k_x,  \delta k_y, \delta k_z )$,  
we have 
$h_x \approx \frac{3}{2} \Delta_{T_{2u}^+} (\delta k_x - \delta k_y) $, 
$h_y \approx  -\sqrt{2} \Delta_{T_{1u} ^-} \delta k_z $ 
and $h_z \approx  \sqrt{3} t (\delta k_x + \delta k_y) $. 
The quasiparticles near the gapless points are thus Weyl fermions~\cite{2011_Leon_PRL,2011_Ashvin_Weyl_PRB} with highly anisotropic velocities. Besides this gapless topological property, it is worth noting that the other two vortex lines that do not cross the Fermi surface are also topologically robust.

In general, we argue that the topology of 3d density waves can be characterized by two integers $(n_1, n_2)$,  $n_1$ being the number of vortex lines that cross the Fermi surface, and $n_2$ the number of other vortex lines that cannot be contracted to one point without touching the Fermi surface.  The topological numbers $n_1$ and $n_2$ thus encode  gapless and gapped topologies, respectively. We emphasize that any two vortex lines connectible by the ${\bf Q}$ vector are equivalent. For gapped density waves, $n_1$ is necessarily zero, and we have a  $Z$ classification, whereas for gapless states, both $n_1$ and $n_2$ can be finite and the classification is $Z\times Z$. Using this scheme,  the ${T_{1u} ^-} + T_{2u}^+ $ state has topological numbers $(1, 2)$, exhibiting mixed topologies. The density wave topology as constructed is measurable by time-of-flight techniques in atomic experiments (see Methods).

\begin{figure}[htp] 
\includegraphics[angle=0,width=\linewidth]{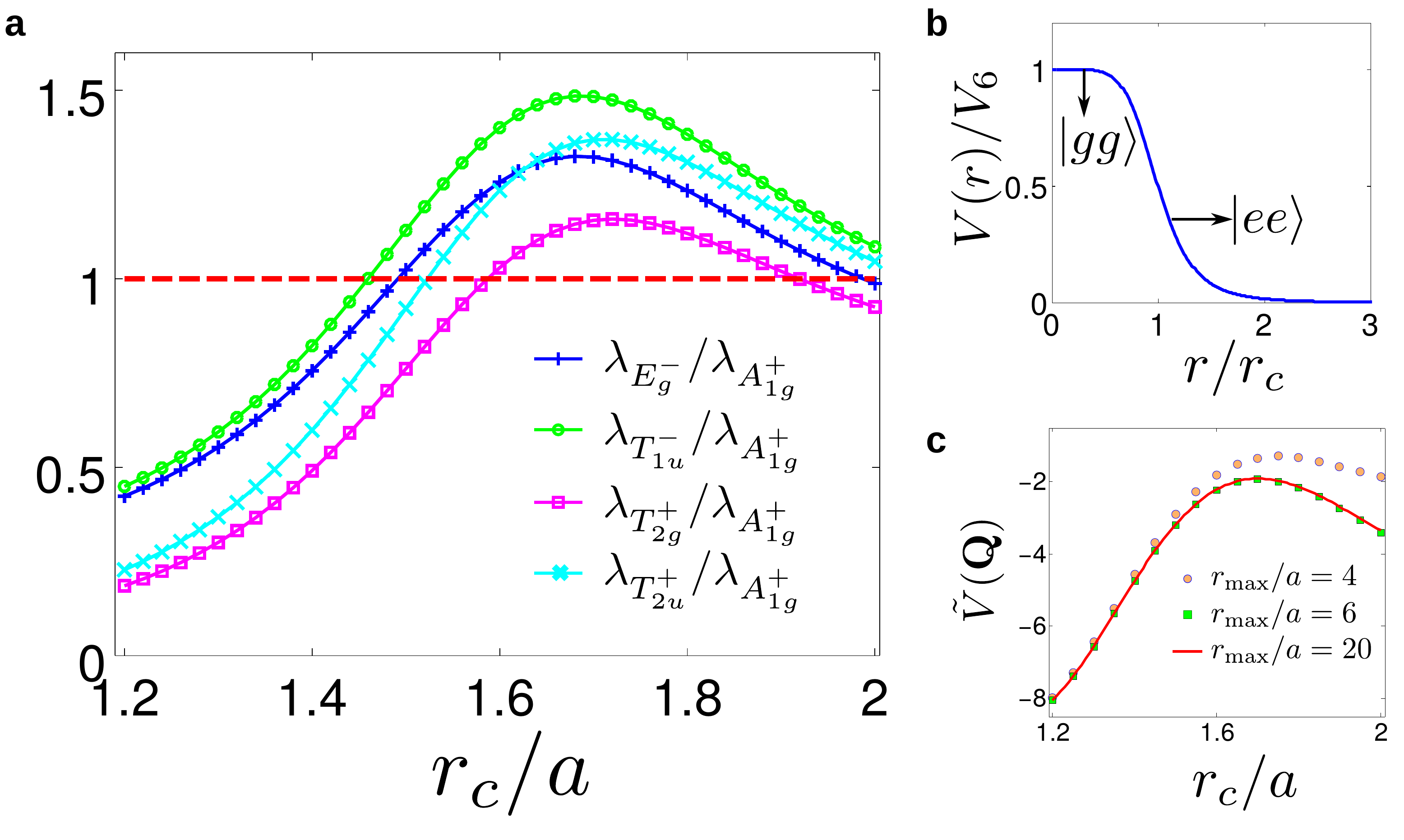}
 \caption{Leading density wave instabilities for Rydberg dressed atomic Fermi gas. {\bf a},  most dominant eigenvalues of the $\gamma$ matrix (see Methods), representing the strengths of particle-hole paring instability of corresponding channels.  
 Increasing $r_c$, the $A_{1g}^+$ channel gets suppressed, and other channels, $T_{1u}^-$, $E_g ^-$, $T_{2g}^+$, and $T_{2u}^+$,  with nontrivial momentum dependence are enhanced.  {\bf b}, step-like interaction form for Rydberg dressed atoms. Two atoms far apart behave as in Rydberg excited states and interact with van-der waals potential,  whereas they become more ground state like at short distance.  {\bf c}, the Fourier transformed interaction $\tilde{V} ({\bf Q})$ [${\bf Q } = (\pi, \pi, \pi)$]. As we increase the cut-off $r_{\rm max}$ (see Methods), $\tilde{V} ({\bf Q})$ converges and the truncation error is already negligible when $r_{\rm max} /r_c = 3$. 
}
\label{fig:quadphasediag} 
\end{figure}

\medskip 
We now discuss the actual density waves supported by Rydberg dressed atoms. With infinitesimal repulsive interaction,  the relative strengths of instabilities in different channels are fully determined by the non-locality strength $r_c/a$. The leading instabilities, as quantified by the corresponding eigenvalues of the $\gamma$ matrix, are shown in Fig.~\ref{fig:quadphasediag}. When $r_c$ is small, the interaction is essentially nearest neighbor, for which the $\gamma$ matrix is  mainly determined by the momentum independent part of interaction $\tilde{V}({\bf Q})$ (the Fourier transform of the interaction at ${\bf Q}$). This makes the trivial $A_{1g}^+$ density wave  completely dominant over all other density wave states. Increasing $r_c$, the longer-ranged part of the interaction becomes more important, which suppresses $\tilde{V}({\bf Q})$ (see Fig.~\ref{fig:quadphasediag}{\bf c}) and consequently the $A_{1g}^+$ density wave. Simultaneously, other density wave channels with non-trivial momentum dependence get enhanced. The dominant instabilities are $T_{1u}^-$, $E_g^-$, $T_{2g} ^+$ and $T_{2u}^+$, and we have a transition from $A_{1g} ^+$ to $T_{1u}^-$ around $r_c/a = 1.46 $. The phase diagram  based on the instability analysis from RG is suggestive, and is rigorous in leading order. However, considering finite interactions, non-linearity  may lead to significant physical effects. In particular, when $r_c$ reaches the scale of lattice spacing, the instability strengths in different channels are quasi-degenerate, which we attribute to the ``step-like" feature of the Rydberg dressed interaction (see Supplementary Information). Non-linear effects must then be taken into account to determine the actual phase diagram.

To incorporate nonlinearity, we numerically simulate a system of size $64\times 64 \times 64$ with periodic boundary condition using self-consistent methods (see Methods). The solution for the ground state  density wave order $\Delta_{\bf k}$ is expanded in terms of basis functions 
\be 
\Delta _{\bf k} = \sum_{\alpha, \beta} \Delta _{\alpha, \beta} \phi_{\alpha, \beta} ({\bf k}), 
\label{eq:DeltaExpansion}
\ee 
where $\phi_{\alpha, \beta} ({\bf k}) $ are the basis functions from symmetry classification (see Supplementary  Table~\tabOhclass), with ${\alpha}$ labeling different classes and ${\beta}$ different functions within one class. 
In the self-consistent theory including non-linear effects, spontaneous symmetry breaking could occur and superpositions of density waves from different classes are allowed, subject to the constraint $\Delta_{\bf k} = \Delta_{ {\bf k} + {\bf Q} } ^*$.  The coefficients $\Delta_{\alpha, \beta}$ are required to be real. 

\begin{figure}[tph] 
\includegraphics[angle=0,width=\linewidth]{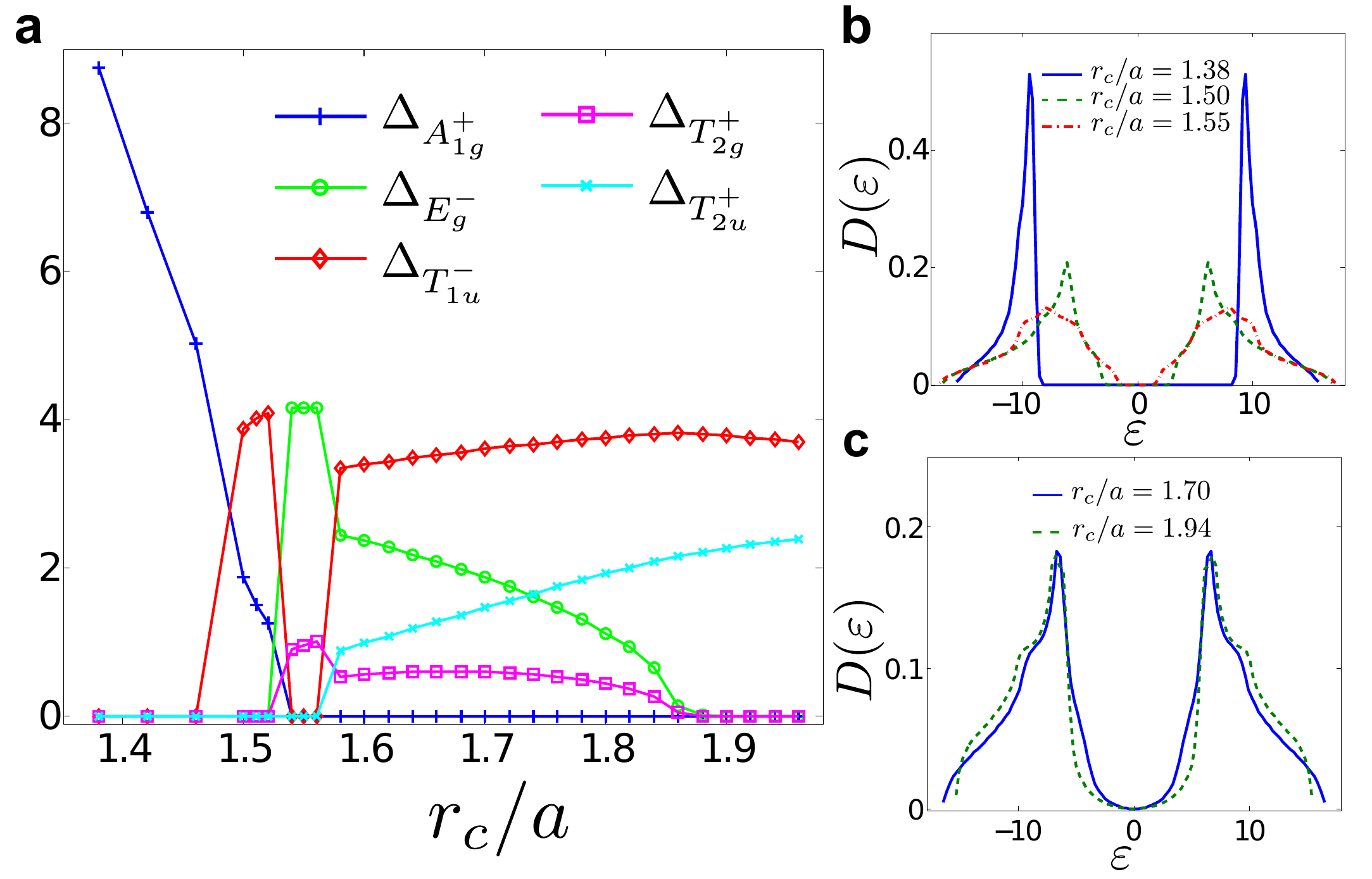}
 \caption{Phase diagram for Rydberg dressed atomic Fermi gas. 
 In this plot, the interaction strength is fixed to be $V_6/t = 4$. 
 To quantify contributions from different classes, we introduce 
$
\Delta_{\alpha} = \sqrt{ \sum_\beta {|\Delta_{\alpha, \beta}|^2} },  
$ 
(see equation~\eqref{eq:DeltaExpansion}). 
 {\bf a} shows the phase diagram with varying Condon radius $r_c$. As we increase $r_c$, 
 we get a sequence of density wave phases, $A_{1g}^+$, $A_{1g}^+ + T_{1u}^-$, $T_{2g}^+ + E_g^-$,  
 $E_g ^{-} + T_{1u}^- + T_{2g} ^+ + T_{2u}^+$ and $T_{1u}^- + T_{2u} ^+$. The first three are fully gapped 
 and the last two are gapless.  
The phase transition between the two gapless phases is found to be second order. The  corresponding broken symmetry  is space inversion.  The phase transitions among the gapped phases are first order. 
 {\bf b} and {\bf c} show the density of states for gapped and gapless phases, respectively. 
}
\label{fig:phasediag} 
\end{figure}

In our self-consistent calculations, spontaneous symmetry breaking of $O_h \times {\cal T}$ is found for a large range of $r_c$ and the symmetry-broken ground states support various superposed density waves, yielding a rich phase diagram (Fig.~\ref{fig:phasediag}). Surprisingly all the superposed states are found to break ${\cal T}$ symmetry and consequently the time-reversal symmetry due to the resultant complex structure in $\Delta_{\bf k}$. The topological  gapless density wave $T_{1u}^- + T_{2u} ^+$ is indeed an allowed solution as argued earlier. The other gapless phase that occupies a large region in the phase diagram is $E_g ^{-} + T_{1u}^- + T_{2g} ^+ + T_{2u}^+$.  In both gapless phases the quasiparticle density of states has  a ``soft-gap" feature (Fig.~\ref{fig:phasediag}c), a signature of the emergent topological Weyl fermions.  In addition, we find  three gapped phases, $A_{1g}^+$, $A_{1g}^+ + T_{1u}^-$ and $E_g ^- +T_{2g}^+$, whose quasiparticle density of states has a ``hard-gap" (Fig.~\ref{fig:phasediag}b).  In atomic experiments, the quasiparticle spectrum and density of states can be directly measured by radio-frequency spectroscopy~\cite{2003_Ketterle_RF_Science,2003_Jin_RF_PRL}.

We now discuss topological properties of the density waves  in the phase diagram. 
In  $E_g ^{-} + T_{1u}^- + T_{2g} ^+ + T_{2u}^+$ state, as compared to $T_{1u}^- + T_{2u} ^+$, the density wave gets additional contributions. The vortex lines are  no longer straight, but otherwise  have the same topology as shown in Fig.~\ref{fig:topology}a. These two gapless density waves are then topologically equivalent, having the same topological numbers $(1,2)$ in our classification scheme. 
In the $ E_g^- + T_{2g}^+ $ state which is gapped, 
the vortex line (with vorticity $2$) is located along one axis (Fig~\ref{fig:topology}b), assumed to be the $k_z$ axis in spontaneous symmetry breaking. This state has topological numbers $(0,2)$. Actually in this state, the quasiparticle Hamiltonian $H_{\rm BdG} ({\bf k})$, with arbitrary fixed $k_z$,  has finite Chern number $2$. Such a state thus exhibits 3d quantum  anomalous Hall effect, featuring chiral surface modes, which mediate highly anisotropic transport properties---dissipationless in $x$ or $y$ direction but diffusive in the $z$ direction~\cite{1996_Balents_3DQHE_PRL,2001_Koshino_3DQHE_PRL}. 
In the other gapped state,  $A_{1g}^+ + T_{1u}^-$, the vortex line is shown in Fig.~\ref{fig:topology}c, but such a vortex line is unstable. 
The $A_{1g}^+ + T_{1u}^-$ state thus has trivial topological numbers $(0,0)$, topologically equivalent to $A_{1g}^+$. 

We expect that our proposed topological density waves to be generic for Rydberg dressed atomic Fermi gases even if the interaction potential is not precisely of the specific $r^6$ form. The key qualitative ingredient is the step-like feature in the interaction. To confirm this, we also have carried out calculations for Rydberg $p$-wave dressed atoms~\cite{2014_Zoller_arXiv}, and  similar topological states are indeed found (see Supplementary Information). We therefore believe that our prediction for novel collective topological density waves should apply generically to dressed Rydberg optical lattices.

\npsection{Methods}
\npsection{Density wave instability under infinitesimal repulsion.} 
The single-particle energy dispersion for the 3d cubic lattice with $O_h$ symmetry is 
$
\epsilon_{\bf k} = -2t \left( \cos k_x + \cos k_y + \cos k_z  \right).  
$ 
Having non-local density-density interaction with potential $V ( {\bf r}- {\bf r}')$, the scattering vertex among single-particle modes 
${\bf k}_1$, ${\bf k}_2$, ${\bf k}_3$ and ${\bf k}_4$ is 
\bea 
&&  \Gamma ({\bf k}_1,{\bf k}_2,{\bf k}_3,{\bf k}_4) 
 = \textstyle  \frac{1}{2} \left\{ \textstyle  \tilde{V} ({\bf k}_2 -{\bf k}_3) + \tilde{V} ({\bf k}_1 - {\bf k}_4) \right. \nn \\
&& 	\textstyle \left.  - \tilde{V} ({\bf k}_2 - {\bf k}_4) - \tilde{V} ({\bf k}_1 - {\bf k}_3)   \right\},  \nn 
\eea 
with $\tilde{V}({\bf q})$ the Fourier transform of $V({\bf r})$.  
At half filling, from the Fermi surface nesting, repulsive interactions cause instabilities in density wave channels. Such instabilities can be tracked by renormalization group (RG) flow of the effective couplings near the Fermi surface, 
$$
g _{ {\bf k}_f, {\bf k}'_f } \equiv \Gamma ({\bf k}' _f + {\bf Q}, {\bf k}_f , {\bf k}'_f, {\bf k} _f+ {\bf Q} ), 
$$ 
The  RG equation at one-loop level~\cite{1994_Shankar_RMP,1994_Weinberg_NPB} reads 
\be
-\Lambda \partial_\Lambda \gamma _{{\bf k}_f, {\bf k}'_f} = 
\int \frac{d ^ 2 {\bf q}_f }{(2\pi)^2} \gamma_{{\bf k}_f, {\bf q}_f } \gamma_{{\bf q}_f , {\bf k}'_f}, \nn 
\ee 
with $\gamma$ a rescaled interaction matrix, 
$$ 
\gamma _{{\bf k}_f, {\bf k}'_f} = \left[ 2\pi v_f ({\bf k}_f)  \right]^{-1/2}  
							g_{{\bf k}_f, {\bf k}'_f}  
					       \left[ 2\pi v_f ({\bf k}'_f) \right] ^{-1/2}, 
$$ 
where $v_f ({\bf k}_f)$ is the Fermi velocity.  In the RG flow, the largest eigenvalue of the $\gamma$ matrix diverges most quickly as approaching the low energy limit. The magnitudes of eigenvalues quantify the relative strength of instabilities in different channels. The symmetry group of $\gamma$ is $O_h\times{\cal T}$, with ${\cal T}$ a two-element group involving the transformation $T({\bf k})= {\bf k } + {\bf Q}$.

\npsection{Non-linear effects and self-consistent theory.} 
The variational state we choose to describe density wave order is~\cite{2000_Nayak_PRB} 
$$ 
\textstyle |\Psi \rangle = \prod_{\bf k} ' \left[ u_{\bf k} \psi ^\dag  ({\bf k}) + v_{\bf k} \psi  ^\dag ({{\bf k} + {\bf Q} })  \right] |\Omega\rangle, 
$$  
with $|\Omega\rangle$ the vacuum, and ${\bf k}$ running over one half of the Brillouin zone, 
e.g., $k_z \ge 0$. The self-energy and density wave order are given by 
\bea 
 && \Sigma_{\bf k} = \frac{1}{2} 
\int _{\bf q} 
\left[ \tilde{V} ({\bf k} -{\bf q} + {\bf Q} ) - \tilde{V} ({\bf k} - {\bf q} ) \right] n _{\bf q} \nn 
\\ 
 && \Delta_{\bf k} =  
	\int _{\bf q} 
	 \left[ \tilde{V} ({\bf Q} ) - \tilde{V} ({\bf k} - {\bf q}) \right]  \rho_{\bf q} , \nn 
\eea
with $\int _{\bf q} \equiv \int \frac{d^3 {\bf q}}{(2\pi)^3} $, 
$n_{\bf q} =  \langle \psi^\dag ( {\bf q})   \psi ({\bf q}) \rangle $, and  
$\rho_{\bf q} = \langle \psi ^\dag ({{\bf q} + {\bf Q}} ) \psi ({\bf q}) \rangle$ . 
The self-consistent equations are obtained to be 
\bea 
&& \Delta_{\bf k}  = \frac{1}{2} 
 \int _{\bf q} \left[ \tilde{V} ({\bf k} - {\bf q}) - \tilde{V} ({\bf Q})  \right] 
	{\Delta_{\bf q} } /{ |\varepsilon_{\bf q}| },  \nn \\ 
&& \Sigma_{\bf k} = \frac{1}{2} 
  \int _{\bf q} \tilde{V} ({\bf k}- {\bf q}) 
	\left[ \epsilon_{\bf q} + \Sigma_{\bf q} \right] / |\varepsilon_{\bf q}|, \nn
\eea 
with quasiparticle spectra $\varepsilon_{\bf q} = \pm \sqrt{(\epsilon_{\bf q} + \Sigma_{\bf q} )^2 + |\Delta_{\bf q}| ^2}$. 
With moderate interaction strength, self-energy corrections could be significant, but the modification to the Fermi surface is reasonably weak (see Supplementary Figure $1$). In numerics, we introduce a long-range cutoff $r_{\rm max}$, so that $V(|{\bf r}|>r_{\rm max})$ is set to be zero. For all numerical results presented in the main text, $r_{\rm max}$ is set to be $6\times a$ and varied to ensure that results do not change qualitatively.

\npsection{Experimental signature of density wave topology.} 
In experiments, the momentum dependence of the density wave  order can be extracted from time-of-flight measurements.  Using mean field theory, it is straightforward to show that 
\be 
|\rho_{\bf k}|^2= \langle n_{\bf k} \rangle \langle n_{{\bf k} + {\bf Q}} \rangle
 - \langle n_{\bf k} n_{{\bf k} + {\bf Q}} \rangle. 
\ee 
At finite temperature, to extract $|\rho_{\bf k}|^2$, it is necessary to measure both  the product $\langle n_{\bf k} \rangle \langle n_{{\bf k} + {\bf Q}} \rangle$ and the correlation $\langle n_{\bf k} n_{{\bf k} + {\bf Q}} \rangle$. At half filling, the latter originates purely from thermal fluctuations.  The former is dominant when the temperature is not too high as compared to the quasiparticle energy gap (or the ``soft-gap" for gapless states).  This implies that it should be experimentally straightforward to get a precise measurement of $|\rho_{\bf k}|^2$. In the zero temperature limit, the density wave order is completely determined by the product $|\rho_{\bf k}|^2 = \langle n_{\bf k} \rangle \langle n_{{\bf k} + {\bf Q}} \rangle$.  Moreover  from the relation, 
\be 
\rho_{\bf k} = -\frac{1}{2} \frac{\Delta_{\bf k}} {\varepsilon_{\bf k}} \left[ n_F (\varepsilon_{\bf k}) - n_F (\varepsilon_{{\bf k } +{\bf Q}}) \right] 
\ee 
the vortex lines of $\Delta_{\bf k}$ immediately follow by tracking the vanishing points of $|\rho_{\bf k}|$. 
With vortex lines located in the 3d momentum space, the topological numbers of density waves  can be easily extracted.

\npsection{Acknowledgements} 

X.L. acknowledges helpful discussions with S. Sachdev, A. Gorshkov, Bo Liu and Xin Liu.  
This work is supported by JQI-NSF-PFC and ARO-Atomtronics-MURI.

\npsection{Author contributions} 

XL and SDS conceived the problem through discussions.  XL carried out the calculations.  Both contributed to the preparation of the manuscript.

\bibliography{references}

\begin{thebibliography}{10}
\expandafter\ifx\csname url\endcsname\relax
  \def\url#1{\texttt{#1}}\fi
\expandafter\ifx\csname urlprefix\endcsname\relax\def\urlprefix{URL }\fi
\providecommand{\bibinfo}[2]{#2}
\providecommand{\eprint}[2][]{\url{#2}}

\bibitem{2009_Low_Rydberg_PRA}
\bibinfo{author}{L\"ow, R.} \emph{et~al.}
\newblock \bibinfo{title}{Universal scaling in a strongly interacting rydberg
  gas}.
\newblock \emph{\bibinfo{journal}{Phys. Rev. A}} \textbf{\bibinfo{volume}{80}},
  \bibinfo{pages}{033422} (\bibinfo{year}{2009}).

\bibitem{2012_Bloch_Rydberg_Nature}
\bibinfo{author}{Schausz, P.} \emph{et~al.}
\newblock \bibinfo{title}{Observation of spatially ordered structures in a
  two-dimensional rydberg gas}.
\newblock \emph{\bibinfo{journal}{Nature}} \textbf{\bibinfo{volume}{491}},
  \bibinfo{pages}{87--91} (\bibinfo{year}{2012}).

\bibitem{2014_Bloch_Rydberg_arXiv}
\bibinfo{author}{{Schau{\ss}}, P.} \emph{et~al.}
\newblock \bibinfo{title}{{Dynamical crystallization in a low-dimensional
  Rydberg gas}}.
\newblock \emph{\bibinfo{journal}{ArXiv e-prints}}  (\bibinfo{year}{2014}).
\newblock \eprint{1404.0980}.

\bibitem{2014_Low_Rydberg_News}
\bibinfo{author}{Low, R.}
\newblock \bibinfo{title}{Rydberg atoms: Two to tango}.
\newblock \emph{\bibinfo{journal}{Nat Phys}} \textbf{\bibinfo{volume}{10}},
  \bibinfo{pages}{901--902} (\bibinfo{year}{2014}).

\bibitem{2014_Browaeys_NatPhys}
\bibinfo{author}{Ravets, S.} \emph{et~al.}
\newblock \bibinfo{title}{Coherent dipole-dipole coupling between two single
  rydberg atoms at an electrically-tuned forster resonance}.
\newblock \emph{\bibinfo{journal}{Nat Phys}} \textbf{\bibinfo{volume}{10}},
  \bibinfo{pages}{914--917} (\bibinfo{year}{2014}).

\bibitem{2009_Kitaev_arXiv}
\bibinfo{author}{{Kitaev}, A.}
\newblock \bibinfo{title}{{Periodic table for topological insulators and
  superconductors}} \textbf{\bibinfo{volume}{1134}}, \bibinfo{pages}{22--30}
  (\bibinfo{year}{2009}).
\newblock \eprint{0901.2686}.

\bibitem{2010_Ryu_NJP}
\bibinfo{author}{Ryu, S.}, \bibinfo{author}{Schnyder, A.~P.},
  \bibinfo{author}{Furusaki, A.} \& \bibinfo{author}{Ludwig, A. W.~W.}
\newblock \bibinfo{title}{Topological insulators and superconductors: tenfold
  way and dimensional hierarchy}.
\newblock \emph{\bibinfo{journal}{New Journal of Physics}}
  \textbf{\bibinfo{volume}{12}}, \bibinfo{pages}{065010}
  (\bibinfo{year}{2010}).

\bibitem{1988_Haldane_QHE}
\bibinfo{author}{Haldane, F.}
\newblock \bibinfo{title}{Model for a quantum hall effect without landau
  levels: Condensed-matter realization of the "parity anomaly"}.
\newblock \emph{\bibinfo{journal}{Phys. Rev. Lett.}}
  \textbf{\bibinfo{volume}{61}}, \bibinfo{pages}{2015--2018}
  (\bibinfo{year}{1988}).

\bibitem{2013_Xue_QAHE}
\bibinfo{author}{Chang, C.-Z.} \emph{et~al.}
\newblock \bibinfo{title}{Experimental observation of the quantum anomalous
  hall effect in a magnetic topological insulator}.
\newblock \emph{\bibinfo{journal}{Science}} \textbf{\bibinfo{volume}{340}},
  \bibinfo{pages}{167--170} (\bibinfo{year}{2013}).

\bibitem{2007_Fu_PRL}
\bibinfo{author}{Fu, L.}, \bibinfo{author}{Kane, C.} \& \bibinfo{author}{Mele,
  E.}
\newblock \bibinfo{title}{Topological insulators in three dimensions}.
\newblock \emph{\bibinfo{journal}{Phys. Rev. Lett.}}
  \textbf{\bibinfo{volume}{98}}, \bibinfo{pages}{106803}
  (\bibinfo{year}{2007}).

\bibitem{2007_Moore_PRB}
\bibinfo{author}{Moore, J.} \& \bibinfo{author}{Balents, L.}
\newblock \bibinfo{title}{Topological invariants of time-reversal-invariant
  band structures}.
\newblock \emph{\bibinfo{journal}{Phys. Rev. B}} \textbf{\bibinfo{volume}{75}},
  \bibinfo{pages}{121306} (\bibinfo{year}{2007}).

\bibitem{2008_Qi_Zhang_PRB}
\bibinfo{author}{Qi, X.-L.}, \bibinfo{author}{Hughes, T.} \&
  \bibinfo{author}{Zhang, S.-C.}
\newblock \bibinfo{title}{Topological field theory of time-reversal invariant
  insulators}.
\newblock \emph{\bibinfo{journal}{Phys. Rev. B}} \textbf{\bibinfo{volume}{78}},
  \bibinfo{pages}{195424} (\bibinfo{year}{2008}).

\bibitem{2009_Essin_Moore_Vanderbilt_TI}
\bibinfo{author}{Essin, A.}, \bibinfo{author}{Moore, J.} \&
  \bibinfo{author}{Vanderbilt, D.}
\newblock \bibinfo{title}{Magnetoelectric polarizability and axion
  electrodynamics in crystalline insulators}.
\newblock \emph{\bibinfo{journal}{Phys. Rev. Lett.}}
  \textbf{\bibinfo{volume}{102}}, \bibinfo{pages}{146805}
  (\bibinfo{year}{2009}).

\bibitem{2011_Leon_PRL}
\bibinfo{author}{Burkov, A.~A.} \& \bibinfo{author}{Balents, L.}
\newblock \bibinfo{title}{Weyl semimetal in a topological insulator
  multilayer}.
\newblock \emph{\bibinfo{journal}{Phys. Rev. Lett.}}
  \textbf{\bibinfo{volume}{107}}, \bibinfo{pages}{127205}
  (\bibinfo{year}{2011}).

\bibitem{2011_Ashvin_Weyl_PRB}
\bibinfo{author}{Wan, X.}, \bibinfo{author}{Turner, A.},
  \bibinfo{author}{Vishwanath, A.} \& \bibinfo{author}{Savrasov, S.}
\newblock \bibinfo{title}{Topological semimetal and fermi-arc surface states in
  the electronic structure of pyrochlore iridates}.
\newblock \emph{\bibinfo{journal}{Phys. Rev. B}} \textbf{\bibinfo{volume}{83}},
  \bibinfo{pages}{205101} (\bibinfo{year}{2011}).

\bibitem{2011_Dai_Weyl_PRL}
\bibinfo{author}{Xu, G.}, \bibinfo{author}{Weng, H.}, \bibinfo{author}{Wang,
  Z.}, \bibinfo{author}{Dai, X.} \& \bibinfo{author}{Fang, Z.}
\newblock \bibinfo{title}{Chern semimetal and the quantized anomalous hall
  effect in ${\mathrm{hgcr}}_{2}{\mathrm{se}}_{4}$}.
\newblock \emph{\bibinfo{journal}{Phys. Rev. Lett.}}
  \textbf{\bibinfo{volume}{107}}, \bibinfo{pages}{186806}
  (\bibinfo{year}{2011}).

\bibitem{2012_Sun_NatPhys}
\bibinfo{author}{Sun, K.}, \bibinfo{author}{Liu, W.~V.},
  \bibinfo{author}{Hemmerich, A.} \& \bibinfo{author}{Das~Sarma, S.}
\newblock \bibinfo{title}{Topological semimetal in a fermionic optical
  lattice}.
\newblock \emph{\bibinfo{journal}{Nat Phys}} \textbf{\bibinfo{volume}{8}},
  \bibinfo{pages}{67--70} (\bibinfo{year}{2012}).

\bibitem{2014_Yang_NatComm}
\bibinfo{author}{Yang, B.-J.} \& \bibinfo{author}{Nagaosa, N.}
\newblock \bibinfo{title}{Classification of stable three-dimensional dirac
  semimetals with nontrivial topology}.
\newblock \emph{\bibinfo{journal}{Nat Commun}} \textbf{\bibinfo{volume}{5}},
  \bibinfo{pages}{4898} (\bibinfo{year}{2014}).

\bibitem{2010_Henkel_PRL}
\bibinfo{author}{Henkel, N.}, \bibinfo{author}{Nath, R.} \&
  \bibinfo{author}{Pohl, T.}
\newblock \bibinfo{title}{Three-dimensional roton excitations and supersolid
  formation in rydberg-excited bose-einstein condensates}.
\newblock \emph{\bibinfo{journal}{Phys. Rev. Lett.}}
  \textbf{\bibinfo{volume}{104}}, \bibinfo{pages}{195302}
  (\bibinfo{year}{2010}).

\bibitem{2014_Zoller_Rydberg_arXiv}
\bibinfo{author}{{Glaetzle}, A.~W.} \emph{et~al.}
\newblock \bibinfo{title}{{Quantum Spin Ice and dimer models with Rydberg
  atoms}}.
\newblock \emph{\bibinfo{journal}{ArXiv e-prints}}  (\bibinfo{year}{2014}).
\newblock \eprint{1404.5326}.

\bibitem{2000_Nayak_PRB}
\bibinfo{author}{Nayak, C.}
\newblock \bibinfo{title}{Density-wave states of nonzero angular momentum}.
\newblock \emph{\bibinfo{journal}{Phys. Rev. B}} \textbf{\bibinfo{volume}{62}},
  \bibinfo{pages}{4880--4889} (\bibinfo{year}{2000}).

\bibitem{2012_Zhao_PRL}
\bibinfo{author}{Bhongale, S.~G.}, \bibinfo{author}{Mathey, L.},
  \bibinfo{author}{Tsai, S.-W.}, \bibinfo{author}{Clark, C.~W.} \&
  \bibinfo{author}{Zhao, E.}
\newblock \bibinfo{title}{Bond order solid of two-dimensional dipolar
  fermions}.
\newblock \emph{\bibinfo{journal}{Phys. Rev. Lett.}}
  \textbf{\bibinfo{volume}{108}}, \bibinfo{pages}{145301}
  (\bibinfo{year}{2012}).

\bibitem{GroupTheory}
\bibinfo{author}{Dresselhaus, M.}, \bibinfo{author}{Dresselhaus, G.} \&
  \bibinfo{author}{Jorio, A.}
\newblock \emph{\bibinfo{title}{Group Theory: Application to the Physics of
  Condensed Matter}} (\bibinfo{publisher}{Springer}, \bibinfo{year}{2008}).

\bibitem{2003_Ketterle_RF_Science}
\bibinfo{author}{Gupta, S.} \emph{et~al.}
\newblock \bibinfo{title}{Radio-frequency spectroscopy of ultracold fermions}.
\newblock \emph{\bibinfo{journal}{Science}} \textbf{\bibinfo{volume}{300}},
  \bibinfo{pages}{1723--1726} (\bibinfo{year}{2003}).

\bibitem{2003_Jin_RF_PRL}
\bibinfo{author}{Regal, C.} \& \bibinfo{author}{Jin, D.}
\newblock \bibinfo{title}{Measurement of positive and negative scattering
  lengths in a fermi gas of atoms}.
\newblock \emph{\bibinfo{journal}{Phys. Rev. Lett.}}
  \textbf{\bibinfo{volume}{90}}, \bibinfo{pages}{230404}
  (\bibinfo{year}{2003}).

\bibitem{1996_Balents_3DQHE_PRL}
\bibinfo{author}{Balents, L.} \& \bibinfo{author}{Fisher, M. P.~A.}
\newblock \bibinfo{title}{Chiral surface states in the bulk quantum hall
  effect}.
\newblock \emph{\bibinfo{journal}{Phys. Rev. Lett.}}
  \textbf{\bibinfo{volume}{76}}, \bibinfo{pages}{2782--2785}
  (\bibinfo{year}{1996}).

\bibitem{2001_Koshino_3DQHE_PRL}
\bibinfo{author}{Koshino, M.}, \bibinfo{author}{Aoki, H.},
  \bibinfo{author}{Kuroki, K.}, \bibinfo{author}{Kagoshima, S.} \&
  \bibinfo{author}{Osada, T.}
\newblock \bibinfo{title}{Hofstadter butterfly and integer quantum hall effect
  in three dimensions}.
\newblock \emph{\bibinfo{journal}{Phys. Rev. Lett.}}
  \textbf{\bibinfo{volume}{86}}, \bibinfo{pages}{1062--1065}
  (\bibinfo{year}{2001}).

\bibitem{2014_Zoller_arXiv}
\bibinfo{author}{{Glaetzle}, A.~W.} \emph{et~al.}
\newblock \bibinfo{title}{{Quantum Spin Ice and dimer models with Rydberg
  atoms}}.
\newblock \emph{\bibinfo{journal}{ArXiv e-prints}}  (\bibinfo{year}{2014}).
\newblock \eprint{1404.5326}.

\bibitem{1994_Shankar_RMP}
\bibinfo{author}{Shankar, R.}
\newblock \bibinfo{title}{Renormalization-group approach to interacting
  fermions}.
\newblock \emph{\bibinfo{journal}{Rev. Mod. Phys.}}
  \textbf{\bibinfo{volume}{66}}, \bibinfo{pages}{129--192}
  (\bibinfo{year}{1994}).

\bibitem{1994_Weinberg_NPB}
\bibinfo{author}{Weinberg, S.}
\newblock \bibinfo{title}{Effective action and renormalization group flow of
  anisotropic superconductors}.
\newblock \emph{\bibinfo{journal}{Nuclear Physics B}}
  \textbf{\bibinfo{volume}{413}}, \bibinfo{pages}{567 -- 578}
  (\bibinfo{year}{1994}).

\bibitem{2010_Raghu_RepSC_PRB}
\bibinfo{author}{Raghu, S.}, \bibinfo{author}{Kivelson, S.~A.} \&
  \bibinfo{author}{Scalapino, D.~J.}
\newblock \bibinfo{title}{Superconductivity in the repulsive hubbard model: An
  asymptotically exact weak-coupling solution}.
\newblock \emph{\bibinfo{journal}{Phys. Rev. B}} \textbf{\bibinfo{volume}{81}},
  \bibinfo{pages}{224505} (\bibinfo{year}{2010}).

\end{thebibliography}
\bibliographystyle{naturemag}

\onecolumngrid

\newpage

\renewcommand{\theequation}{S\arabic{equation}}
\renewcommand{\tablename}{{\bf Supplementary Table}}
\renewcommand{\figurename}{{\bf Supplementary Figure}}
\setcounter{equation}{0}  
\setcounter{figure}{0}  
\setcounter{table}{0} 
\renewcommand{\thetable}{{\bf \arabic{table}} }
\renewcommand{\thefigure}{{\bf \arabic{figure}}}
\renewcommand{\thesection}{S-{{\bf \arabic{section}}}} 
\section*{\Large\bf Supplementary Information}
\begin{center}
{\large \bf \papertitle} 
\end{center}

\def\scphasediag{3 }

\section{\bf A consistent theory for density wave order} 
\label{sec:basictheory} 
The model Hamiltonian to describe a Rydberg dressed atomic Fermi gas in a three dimensional (3d) cubic optical lattice is $H = H_0 + H_{\rm int}$ 
with 
\bea
H_{0} &=& \sum_{\bf r, \alpha}\left[ -t_{\alpha } c_{\bf r} ^\dag c_{ {\bf r} + a \hat{e}_\alpha } + h.c.  \right],  \nn \\ 
H_{\rm int} &=& \frac{1}{2}  \sum_{{\bf r}, {\bf r}'} V ( {\bf r}- {\bf r}') c^\dag_{\bf r} c^\dag_{ {\bf r}'} c_{{\bf r}'} c_{\bf r}.  
\eea
Here $a$ is the lattice constant,  $\alpha$ index the three directions $x$, $y$ and $z$, and $c_{\bf r} $  the fermionic annihilation operator positioned at ${\bf r}$. 
The atomic interaction potential, assuming the Rydberg state is $s$-wave, takes an isotropic form, 
$
V({\bf r}) = {V_6} /[{1+(|{\bf r}|/r_c)^6}], 
$ 
for $r \neq 0$. For fermionic atoms considered here, 
$V(0)$ is set to be zero.

To describe a density wave ground state,  we take a variational state 
\be 
\textstyle |\Psi \rangle = \prod_{\bf k} ' \left[ u_{\bf k} c_{\bf k} ^\dag + v_{\bf k} c_{{\bf k} + {\bf Q} } ^\dag \right] |\Omega\rangle. 
\ee 
Here $|\Omega\rangle$ is the vacuum, and we have a normalization condition 
$|u_{\bf k}|^2 + |v_{\bf k}|^2=1$. In $\prod_{\bf k} '$, ${\bf k}$ runs over one half of the Brillouin zone, 
say $k_z \ge 0$. We can define $u_{\bf k}$ and $v_{\bf k}$ in the whole Brillouin zone by the relation 
$u_{{\bf k} + {\bf Q}} = v_{\bf k}$, $v_{{\bf k} + {\bf Q}} = u_{\bf k}$. Half filling is already enforced within this variational ansatz. 
The variational energy cost $E = \langle \Psi | H_0 + H_{\rm int} |\Psi \rangle$  is  obtained to be 
\bea 
E 
 &=& -\frac{1}{2} \int \frac{d^3  {\bf k}}{(2\pi)^3}  
 	\sqrt{(\epsilon_{\bf k} + \Sigma_{\bf k} )^2 + |\Delta_{\bf k}| ^2 }  \nn \\
& &+ \frac{1}{2} \int  \frac{d ^3 {\bf k}}{(2\pi)^3}  \frac{d^3 {\bf k}'}{(2\pi)^3}   
	\Delta ^* _{\bf k} \tilde{V}_{\bf Q}  ^{-1} ({\bf k} - {\bf k}') \Delta _{{\bf k}'}, 
\eea 
with 
$\tilde{V}_{\bf Q} ({\bf k} - {\bf k} ' ) = \tilde{V} ({\bf k} - {\bf k}') - \tilde{V} ({\bf Q}) $. 
Minimizing the energy, we get the Bogoliubov-de Gennes (BdG) equation 
\be 
H_{\rm BdG} ({\bf k}) 
\left[ 
	\begin{array}{c}
	u_{\bf k} \\ 
	v_{\bf k}  
	\end{array} 
\right]  
= \varepsilon_{\bf k} 
\left[ 
	\begin{array}{c}
	u_{\bf k} \\ 
	v_{\bf k}  
	\end{array} 
\right] , 
\ee 
with 
\be 
H_{\rm BdG} ({\bf k}) 
= 
\left[ 
\begin{array}{cc} 
\epsilon_{\bf k} +  \Sigma_{\bf k} & \Delta_{\bf k} \\ 
\Delta ^*_{\bf k} 	& 	-\epsilon_{\bf k} - \Sigma_{\bf k} 
\end{array} 
\right],  
\ee 
where the self-energy is 
\be
\Sigma_{\bf k} = \frac{1}{2} 
\int \frac{d ^3 {\bf q}}{(2\pi)^3} 
\left[ \tilde{V} ({\bf k} -{\bf q} + {\bf Q} ) - \tilde{V} ({\bf k} - {\bf q} ) \right] | u_{\bf q} | ^2 
\ee 
and the off-diagonal term is the density wave order, 
\be 
\Delta_{\bf k} = 
	\int \frac{d^3 {\bf q}}{(2\pi)^3} 
	 \left[ \tilde{V} ({\bf Q} ) - \tilde{V} ({\bf k} - {\bf q}) \right]  u_{\bf q} v^* _{\bf q}. 
\ee 
The chemical potential has been shifted by  $\frac{1}{2} \tilde{V} (0)$ to compensate Hartree corrections 
to the self-energy. By definition, the density wave order satisfies 
$ 
\Delta_{\bf k} = \Delta ^* _{{\bf k} + {\bf Q}}, 
$  
which is an important difference of particle-hole pairing from particle-particle paring in superconducting states. 
Diagonalizing the BdG Hamiltonian, quasiparticle energies are given as 
$ 
\varepsilon_{\bf k} = \pm \sqrt{ (\epsilon_{\bf k} + \Sigma_{\bf k} )^2 + |\Delta_{\bf k} | ^2 }. 
$  
In variational calculations, we choose the negative branch, and 
the self-consistent equations are then obtained to be 
\bea 
&& \Delta_{\bf k}  = \frac{1}{2} 
 \int \frac{d^3 {\bf q}}{(2\pi)^3} \left[ \tilde{V} ({\bf k} - {\bf q}) - \tilde{V} ({\bf Q})  \right] 
	{\Delta_{\bf q} } /{ |\varepsilon_{\bf q}| },  \nn \\ 
&& \Sigma_{\bf k} = \frac{1}{2} 
  \int \frac{d^3 {\bf q}}{(2\pi)^3} \tilde{V} ({\bf k}- {\bf q}) 
	\left[ \epsilon_{\bf q} + \Sigma_{\bf q} \right] / |\varepsilon_{\bf q}|.  
\label{eq:supscequations} 
\eea

The effective potential for $\Delta_{\bf k}$, defined by  
$U[\Delta_{\bf k}] = \left( E[\Delta_{\bf k}] - E[0] \right)/N_s $, 
reads 
\bea
U[\Delta_{\bf k}] 
 &=& -\frac{1}{2} \int \frac{d^3 {\bf k}}{(2\pi)^3} 
 	\left\{ \sqrt{(\epsilon_{\bf k} + \Sigma_{\bf k} )^2 + |\Delta_{\bf k}| ^2 } 
		- |\epsilon_{\bf k} + \Sigma_{\bf k}| \right\} \nn \\
&& + \frac{1}{2} \int \frac{d^3{\bf k}}{(2\pi)^3 } \frac{d^3 {\bf k}'} {(2\pi)^3 } 
	\Delta ^* _{\bf k} \tilde{V}_{\bf Q}  ^{-1} ({\bf k} - {\bf k}') \Delta _{{\bf k}'}.   
\eea 
With weak interaction, the density wave order $\Delta_{\bf k} $ is restricted to 
the region near the Fermi surface. We thus rewrite the momentum in terms of 
Fermi momentum and  the distance from the Fermi surface as 
\be 
{\bf k} = {\bf k}_f + l e_{\perp}, 
\ee 
where $e_{\perp}$ is a unit vector perpendicular to the Fermi surface. The effective potential  now reads 
\bea
&& U[\Delta_{\bf k} ] = -\frac{1}{2} \int \frac{d^2 {\bf k}_f }{(2\pi)^2 } \int _0 ^ \Lambda \frac{dl}{2\pi} 		
\left[ \sqrt{v_f ^2 ({\bf k}_f) l^2 + |\Delta_{{\bf k}_f} |^2   } -v_f ({\bf k}_f) l \right]  
+ \frac{1}{2} \int \frac{ d^2 {\bf k}_f }{(2\pi)^2} \frac{d^2 {\bf k}_f '}{(2\pi)^2 } 
		\Delta ^* _{{\bf k}_f} g ^{-1} _{ {\bf k}_f  {\bf k}_f '} \Delta _{{\bf k}_f '}, 
\eea
where the coupling matrix $[g]$ is the matrix $[\tilde{V}_{\bf Q} ]$ projected onto the Fermi surface. 
To see the momentum dependence of the $\Delta_{\bf k}$, we expand $U[\Delta_{\bf k}]$ to the second order as,  
\bea
U[\Delta] \approx -\frac{1}{4 \pi} \ln \left( \frac{\Lambda}{\tau } \right) 
		\int \frac{d^2 {\bf k}_f } {(2\pi)^2} \frac{ |\Delta_{{\bf k}_f} |^2  }{v_f ({\bf k}_f) }  
	 +  \frac{1}{2} \int \frac{ d^2 {\bf k}_f }{(2\pi)^2} \frac{d^2 {\bf k}_f '}{(2\pi)^2 } 
	 	\left[ \frac{ \Delta ^* _{{\bf k}_f }}{\sqrt{v_f ({\bf k}_f) }} \right]
			\sqrt{v_f ({\bf k}_f) } g^{-1} _{{\bf k}_f  {\bf k}_f '}\sqrt{ v_f ({\bf k}_f ') } 
			 \left[ \frac{ \Delta _{{\bf k}_f '}}{\sqrt{v_f ({\bf k}_f') }} \right], 
\eea 
where an infrared cutoff $\tau$ is introduced to regularize the perturbative expansion. This scale $\tau$ can be physically identified to be temperature. 
Minimizing the effective potential, 
$ \frac{ \Delta _{{\bf k}_f }}{\sqrt{v_f ({\bf k}_f) }} $ is  an eigenvector of the matrix 
\be 
\sqrt{v_f ({\bf k}_f) } g^{-1} _{{\bf k}_f , {\bf k}_f '}\sqrt{ v_f ({\bf k}_f ') } \equiv \gamma^{-1} _{{\bf k_f} {\bf k_f '}}
\label{eq:gamma}
\ee 
with the minimal eigenvalue, 
or equivalently the eigenvector of 
$\gamma_{{\bf k_f} {\bf k_f}'}$ 
with the maximal eigenvalue $\lambda_{\rm max} $. The corresponding  transition temperature is estimated to be 
\be 
\tau_c = \Lambda \exp ( - 2\pi \lambda_{\rm max} ^{-1}  ),  
\ee 
which is valid for infinitesimal interactions. 

It is worth noting that the couplings $g_{{\bf k}_f {\bf k}_f'} $ actually depend on the cutoff $\Lambda$. 
The effective potential $U[\Delta]$ is, on the other hand, independent of $\Lambda$~\cite{1994_Weinberg_NPB}, i.e., 
$$\Lambda \partial_{\Lambda} U[\Delta_{\bf k}] =0,$$ 
from which the renormalization group $\beta$-function follows 
\be 
-\Lambda \partial_{\Lambda}  g_{{\bf k}_f  {\bf k}_f '} = 
\int \frac{d ^2 {\bf q}_f} {(2\pi)^2}  g_{{\bf k}_f  {\bf q}_f } 
	[2\pi v_f ({\bf q}_f) ]^{-1} g_{{\bf q}_f {\bf k}_f' } .  
\ee 
This renormalization equation can also be derived by momentum shell scheme~\cite{1994_Shankar_RMP}. 
Analyzing the renormalization group flow, the momentum dependence of the density wave order can 
be determined from leading divergent channels in the renormalization group flow, analogous to 
the procedure of extracting dominant channels in unconventional superconductivity~\cite{2010_Raghu_RepSC_PRB}. 
This approach is equivalent to minimizing the effective potential at quadratic order. 

\begin{figure}[htp] 
\includegraphics[angle=0,width=\linewidth]{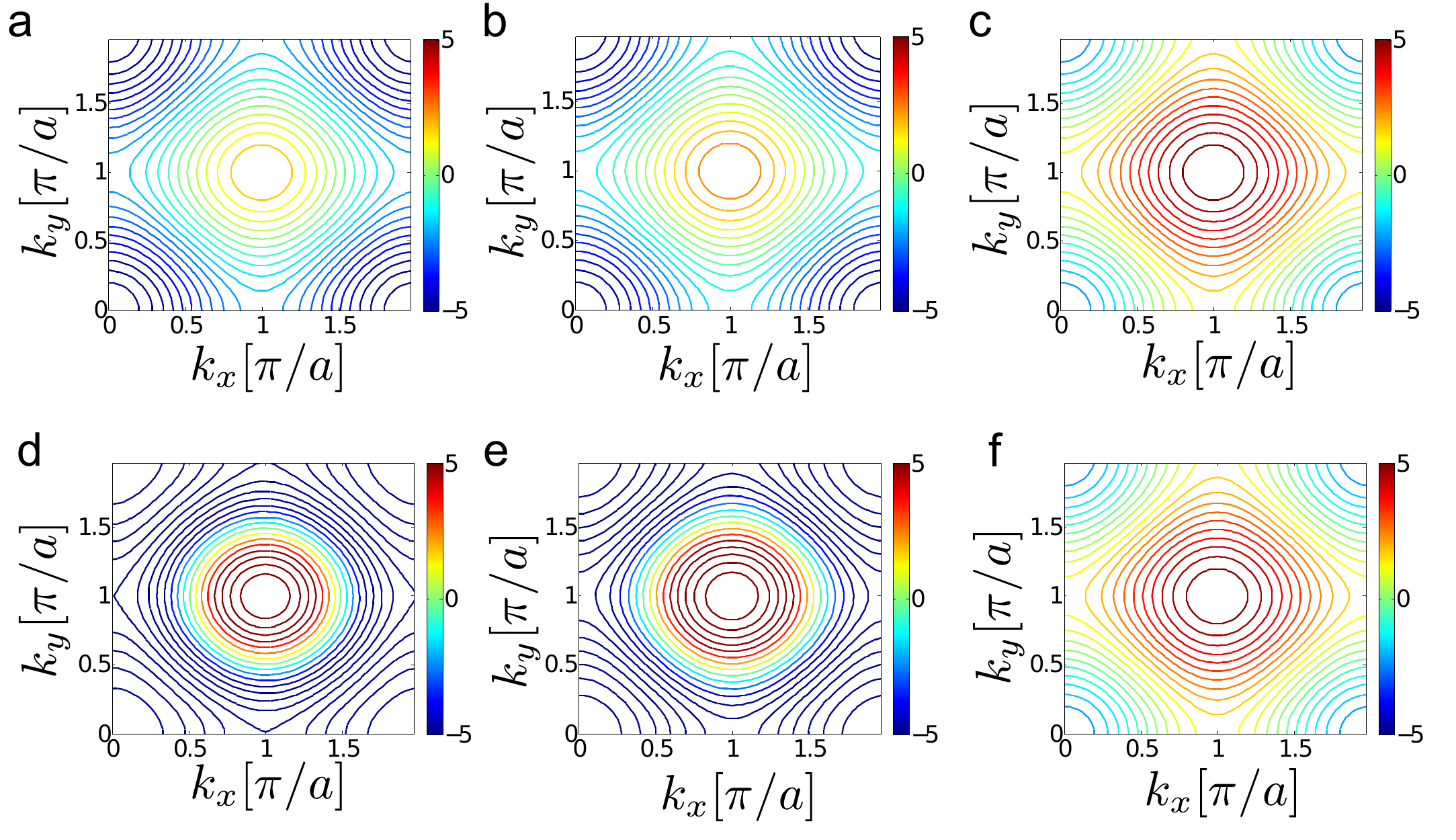}
 \caption{ Illustration of self-energy corrections. In this plot we choose $r_c/a = 1.86$ and $V_6/t = 4$. {\bf a}, {\bf b} and {\bf c} show contour plots of bare dispersion $\epsilon_{\bf k}$ with $k_z a = 0$, $\frac{\pi}{4}$ and $\frac{3 \pi}{4} $, respectively. {\bf d}, {\bf e} and {\bf f} show $\epsilon_{\bf k} + \Sigma_{\bf k} $  for the same set of $k_z$. Although self-energy corrections are significant,  its modification to the Fermi surface (defined by $\epsilon_{\bf k} + \Sigma_{\bf k} =0 $) is weak. The property holds for other choice of parameters as well.  
 }
\label{fig:selfenergy} 
\end{figure}

\section{An approximate criterion for unconventional density waves to dominate} 
As discussed in Section~\ref{sec:basictheory}, the leading density wave instability is determined by the eigenvalue problem of the $\gamma$ matrix (equation~\eqref{eq:gamma}), which in general requires numerics.  Here we would like to derive an approximate criterion for unconventional density waves to occur, and give some intuition why such orders are supported by Rydberg dressed interactions, assuming that the momentum dependence of fermi velocity is negligible.  Under this assumption, the eigenvalue problem of $\gamma$ becomes equivalent to that of  
$$
\tilde{V} ({\bf k} - {\bf k}') - \tilde{V}({\bf Q}) .
$$ 
From Fourier transformation, we know 
\be 
\int \frac{d ^3 {\bf k}'} {(2\pi)^3} [\tilde{V} ({\bf k}- {\bf k}')  - \tilde{V}({\bf Q})  ] e^{-i {\bf k}' \cdot {\bf r}} 
= [V({\bf r}) -  \tilde{V}({\bf Q})\delta_{{\bf r}, 0} ]   e^{- i {\bf k} \cdot {\bf r}}. 
\ee  
The density waves, $\Delta_{\bf k}$, are then given by the plane waves $e^{-i{\bf k}\cdot {\bf r}}$, ${\bf r}$ labeling different solutions. The eigenvalue associated with the trivial density wave with ${\bf r}=0$ is then $-\tilde{V}({\bf Q})$.  For other density waves having momentum dependence,  with ${\bf r} \neq 0$, the eigenvalues are $V({\bf r} ) $. With isotropic interactions, the density wave solutions with the same $|{\bf r}|$ are degenerate. In particular with step-like interactions as for Rydberg dressed atoms, the solutions with $|{\bf r}| <r_c$ are actually all quasi-degenerate. This would strongly amplify non-linear effects, giving rise to possibilities for superposed density waves. 

\begin{figure}[htp] 
\includegraphics[angle=0,width=.6\linewidth]{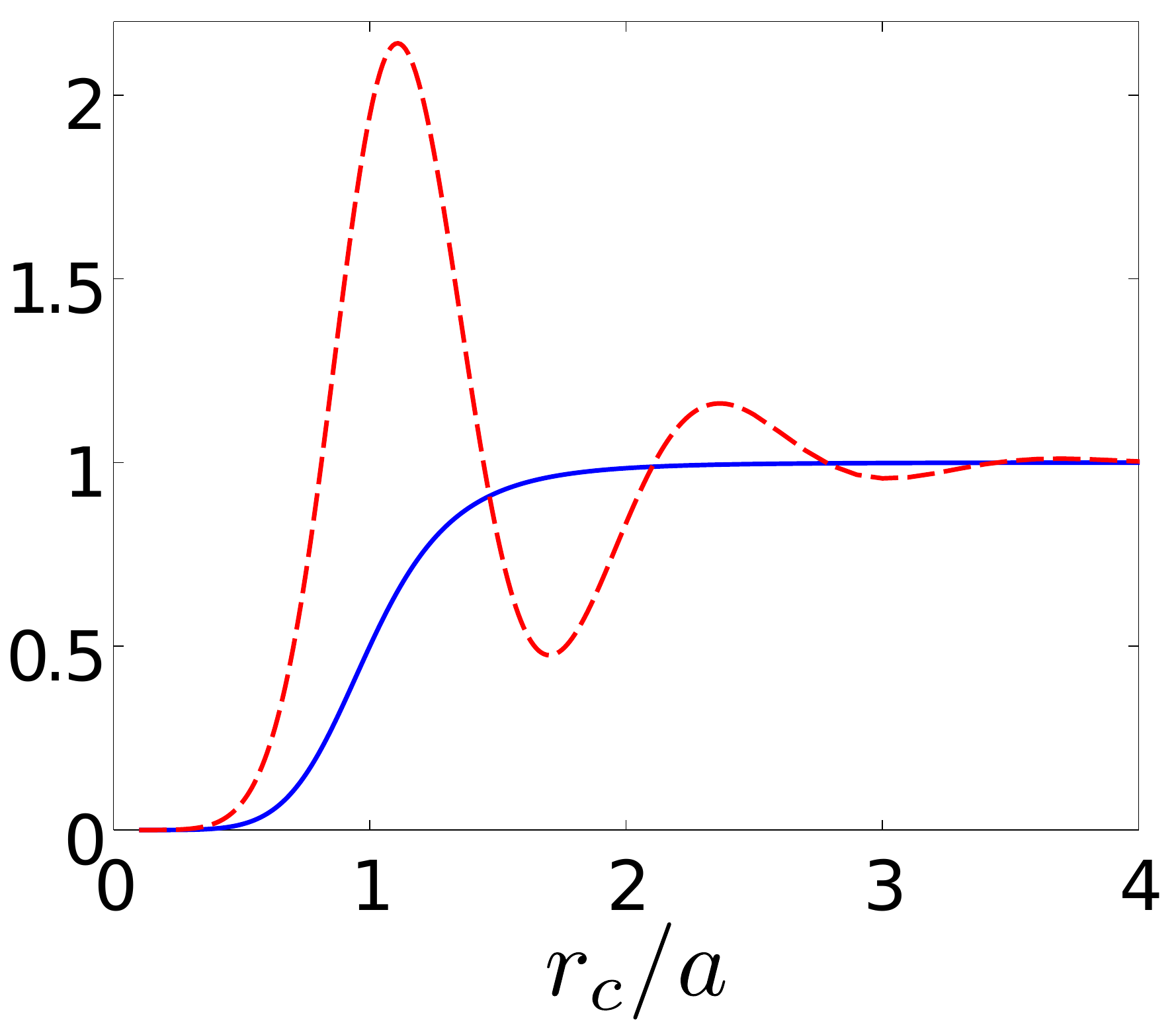}
 \caption{Estimation of relative instability strength towards trivial and non-trivial density waves. 
 The solid and dashed lines correspond to $\frac{1}{1+ (a/r_c)^6}$ and $ -  \sum_{\bf r \neq 0 } \frac{(-1)^{r_x/a + r_y/a + r_z/a} }{1+(|{\bf r}|/r_c)^6}$, respectively, representing instability strengths of momentum independent  and dependent density waves. The latter is significantly dominant in the region,  $r_c/a\in (1.45, 2.0)$. 
}
\label{fig:VQvsVr} 
\end{figure}

The criterion for unconventional density wave to dominate over the trivial one is, 
\be  
\max \{V({\bf r} ) \} > - \tilde{V}({\bf Q}),
\label{eq:generalcriterion} 
\ee 
which implies for the Rydberg interaction that   
\be 
\frac{1}{1+ (a/r_c)^6} > -  \sum_{\bf r \neq 0 } \frac{(-1)^{r_x/a + r_y/a + r_z/a} }{1+(|{\bf r}|/r_c)^6}.  
\ee 
From Supplementary Figure~\ref{fig:VQvsVr}, the parameter range $r_c/a\in (1.45, 2.0)$ should be ideal  to look for unconventional density waves, which is precisely the region we focus on in the main text. For general interactions, the most likely region to look for unconventional density waves can be easily identified by the derived criterion in equation~\eqref{eq:generalcriterion}.

\section{\bf Numerical tricks in solving self-consistent equations} 
In self-consistent iteration (equation~\eqref{eq:supscequations}), a brute force treatment would lead to numerical cost of ${\cal O} (N_s ^2 )$, with $N_s$ the number of lattice sites, which  makes the  numerics too heavy  for large systems, say with $N_s = 64 ^3$.  We simplify the problem by rewriting 
\bea
&&  \Delta_{\bf k}  = \frac{1}{2} 
 \sum_{\bf r} e^{-i {\bf k} \cdot {\bf r}}  V({\bf r}) 
 	\left[ \int \frac{d^3 {\bf q}}{(2\pi)^3}  e^{i{\bf q} \cdot {\bf r} } 
		{\Delta_{\bf q} } /{ |\varepsilon_{\bf q}| }\right]  
	- \frac{1}{2}  \tilde{V} ({\bf Q})  \int \frac{d^3 {\bf q}}{(2\pi)^3}  {\Delta_{\bf q} } /{ |\varepsilon_{\bf q}| }   ,  \nn \\ 
&& \Sigma_{\bf k} = \frac{1}{2} 
\sum_{\bf r} e^{-i {\bf k} \cdot {\bf r}}  V({\bf r}) 
 	\left[ \int \frac{d^3 {\bf q}}{(2\pi)^3}  e^{i{\bf q} \cdot {\bf r} } 
	\left( \epsilon_{\bf q} + \Sigma_{\bf q} \right) / |\varepsilon_{\bf q}| \right] .  
\eea 
Following the bracketing order above in numerical iteration, the cost is greatly reduced to ${\cal O} \left((r_{\rm max }/a)^3 N_s \right)$, ($r_{\rm max}$ is the long-range cutoff, fixed to be $6\times a$ in calculations),  which makes it feasible to simulate  large systems.

\section{\bf Symmetry classification of density waves  } 
In this section we give the classification of density waves according to $O_h \times {\cal T}$ (see Methods in the main text).  The basis functions for density waves with different symmetries can be constructed by a standard projection method in group theory~\cite{GroupTheory}. Classification of density waves and representative irreducible basis functions are shown in Supplementary Table~\ref{tab:classification}. The basis functions are orthonormal. Respecting all symmetries, superpositions of density waves from different symmetry classes are not allowed.

\begin{table*} [htp]
\begin{tabular} {| c | c | c | c | } 
\hline 
$A_{1g} ^+ $		& $A_{1g} ^{-}$		& $A_{1u}^{+}$		& $A_{1u}^{ -}$ \\ \hline 
$1, \cos k_x \cos k_y + \cos k_y \cos k_z + \cos k_z \cos k_x$, 
				&  $i \left( \cos k_x + \cos k_y + \cos k_z \right) $ 
								& \, 				& \, 
\\ \hline \hline 
$A_{2g} ^{+}$		&$A_{2g}^{-} $		& $A_{2u}^{+}$		&$ A_{2u}^-$  \\ \hline 
\, 				& \, 				&\, 				&$\, $
\\ \hline \hline 
$E_g ^+$			&$E_{g} ^-$		&$E_{u}^+$		& $E_{u} ^-$ \\ \hline 
$\left\{ \begin{array}{c} 
	\cos k_z (\cos k_x - \cos k_y)  \\ 
	2 \cos k_x \cos k_y - \cos k_z \left(\cos k_x + \cos k_y\right)
	\end{array} 
	\right.$  			
				&$\left\{ \begin{array}{c}
				 i\left( \cos k_x - \cos k_y \right) \\
				i \left( 2 \cos k_z - \cos k_x - \cos k_y   \right) 
	   			\end{array} 
 				\right. $ 
 								& \, 				& \, 
 \\ \hline \hline 
$T_{1g}^+$ 		&$T_{1g}^-$		&$T_{1u} ^{+}$		& $T_{1u}^-$  \\ \hline 
\, 				& \, 				
								&$ \left\{ \begin{array}{c} 
								\sin k_x \left( \cos k_y + \cos k_z \right)  \\ 
								\sin k_y \left( \cos k_z + \cos k_x \right) \\
								\sin k_z \left( \cos k_x + \cos k_y \right) 
								\end{array} \right. $ 
												& $ \left\{ \begin{array}{c} 
													i \sin k_x \\ 
													i \sin k_y \\
													i \sin k_z 
													\end{array} \right. $ 
\\ \hline  \hline 
$T_{2g} ^+$		&$T_{2g} ^-$		&$T_{2u} ^+$		&$T_{2u}^-$   \\  \hline 
$\left\{ \begin{array}{c} 
	\sin k_x \sin k_y \\ 
	\sin k_y \sin k_z \\
	\sin k_z \sin k_x 
	\end{array} \right. $ 
				&\, 				&$ \left\{ \begin{array}{c}  
									\sin k_x \left( \cos k_y - \cos k_z \right) \\ 
								         \sin k_y \left( \cos k_z - \cos k_x \right) \\
								         \sin k_z \left( \cos k_x - \cos k_y \right) 
								         \end{array} \right. $ 
								         			&\, \\ 
\hline
\end{tabular} 
\caption{Classification of three dimensional density wave orders  according to irreducible representation of the symmetry group $O_{h}\times{\cal T}$. In the labeling of different classes, the superscript $\pm$ tells how the density wave transforms under ${\cal T}$ symmetry. The basis functions representing particle-hole paring of upto next nearest neighboring sites are given. Due to the constraint $\Delta_{\bf k} = \Delta_{{\bf k} + {\bf {Q} } } ^*$, the basis functions in ${\cal T}$ even (odd) classes are real (purely imaginary). The normalization constants for the basis functions are neglected in this table to save writing. 
} 
\label{tab:classification} 
\end{table*}

\section{Topological properties of density waves} 
Details of topological properties of density waves in the self-consistent phase diagram (Fig.~\scphasediag in the main text) are provided in this section. 
In  $E_g ^{-} + T_{1u}^- + T_{2g} ^+ + T_{2u}^+$ state, as compared to $T_{1u}^- + T_{2u} ^+$, 
$\Delta _{\bf k}$ gets an additional 
contribution---$ \sqrt{2} \Delta_{T_{2g}^+ } \sin k_z (\sin  k_x - \sin k_y)   
			+ i   \Delta_{E_g^-}\left[ 2 \cos k_z - \cos k_x - \cos k_y \right] /\sqrt{3}$. 
Treating these additional components, $\Delta_{E_g^-}/\Delta_{T_{1u} ^-} $ and $\Delta_{T_{2g}^+ }/\Delta_{T_{2u}^+} $ as perturbations, we get three vortex lines given by 
$  
\left( 0, l, {\Delta_{E_g^-}}/{(\sqrt{6}\Delta_{T_{1u} ^-}) } \,(\cos l - 1) \right),
$ 
$ 
\left(l, 0,  {\Delta_{E_g^-}}/{(\sqrt{6}\Delta_{T_{1u} ^-} )}\, (\cos l - 1) \right),
$ and 
$ 
\left(l, l, k_z = {2 \Delta_{ E_g^-}}/{(\sqrt{6}\Delta_{T_{1u} ^-} )}\, (\cos l  - 1) \right),
$  
the third of which crosses Fermi surface. The $E_g ^{-} + T_{1u}^- + T_{2g} ^+ + T_{2u}^+$ state is topologically equivalent to  $T_{1u}^- + T_{2u} ^+$, having the same topological numbers $(1,2)$. 

In the $ E_g^- + T_{2g}^+ $ state which is gapped, $\Delta _{\bf k}$ takes a form 
$$ 
\Delta_{\bf k} \approx 2 \Delta_{T_{2g}^+} \sin k_x \sin k_y + i \Delta_{E_g ^- } (\cos k_x - \cos k_y). 
$$ 
The resulting vortex line (with vorticity $2$) is located along the $k_z$ axis (note that the $O_h$ symmetry has been spontaneously broken), thus not crossing the Fermi surface (see Fig.~\ref{fig:topology}b in the main text). This state has topological numbers $(0,2)$ in our classification scheme. 
In the other gapped state,  $A_{1g}^+ + T_{1u}^-$,  we have 
\bea 
\textstyle \Delta_{\bf k} & \approx& \Delta_{A_{1g}^+, 1} - 2/\sqrt{3} \Delta_{A_{1g}^+, 2} 
\sum_{\alpha \neq \alpha'} \cos k_\alpha \cos k_{\alpha'} \nn \\ 
&+& \textstyle  i \sqrt{2} \Delta_{T_{1u}^-} \sin k_z. \nn
\eea 
The vortex line of this state is shown in Fig.~\ref{fig:topology}c in the main text, but it is  unstable because it can be adiabatically contracted to one point without touching the Fermi surface. The $A_{1g}^+ + T_{1u}^-$ state has trivial topological numbers $(0,0)$, topologically equivalent to $A_{1g}^+$.

\section{Density waves with Rydberg $p$-wave dressed atoms}  

In this section we present results for density waves in $p$-wave dressed Rydberg atoms~\cite{2014_Zoller_arXiv}, 
whose interaction takes an anisotropic form 
\be 
V({\bf r}\neq 0 ) = \frac{V_6}{1+(|{\bf r}|/r_c)^6}  \sin ^4 (\theta),  
\ee 
with $\theta$ the polar angle. 
The symmetry group to classify density waves in this system is then $D_{4h} \times {\cal T}$. The symmetry classification is shown in Supplementary Table~\ref{tab:Dhclassification}.

\begin{table*}[htp] 
\begin{tabular} {| c | c | c | c | } 
\hline 
$A_{1g} ^{+}$		& $A_{1g} ^{-}$		& $A_{1u}^{+}$		& $A_{1u}^{ -}$ \\ \hline 
$1$,	$\cos k_x \cos k_y$, $(\cos k_x + \cos k_y ) \cos k_z$ 
			&  $i (\cos k_x + \cos k_y)$, $i \cos k_z$  
						& $( \cos k_x + \cos k_y) \sin k_z$ 
									& $i \sin k_z$ 
\\ \hline \hline 
$E_g ^{+ }$		&$E_g ^-$    		& $E_u ^{+}$		&$E_u ^{-}$ 
\\ \hline 
$\left\{ 
\begin{array}{c}  
 \sin k_x \sin k_z \\ 
 \sin k_y \sin k_z  
\end{array}\right. 
$
			&\,  			&$\left\{ 
						  \begin{array}{cc} 
			     			  \sin k_x \cos k_y, & \sin k_x \cos k_z \\ 
			     			  \cos k_x \sin k_y, & \sin k_y \cos k_z  
						  \end{array} \right. $
									& $\left\{ 
									  \begin{array}{c} 
									   i \sin k_x \\ 
									   i \sin k_y 
									  \end{array}\right. $ 
\\ \hline  \hline 
$B_{1g} ^{+}$ 		& $B_{1g}^{-}$		& $B_{1u } ^+$		& $B_{1u}^-$ 
\\  \hline 
$( \cos k_x - \cos k_y) \cos k_z$ 
			& $i ( \cos k_x -\cos k_y)$ 
						&$(\cos k_x - \cos k_y) \sin k_z$ 
									&\, 
\\ \hline \hline 
$B_{2g} ^{+}$		&$B_{2g} ^-$		& $B_{2u}^+$		&$B_{2u}^-$ 
\\ \hline 
$\sin k_x \sin k_y$	&\, 			&\, 			&\,   
  \\ 
\hline
\end{tabular} 
\caption{Classification of density wave orders for non-local interactions up to next nearest neighbor  according to irreducible representation of the symmetry group $D_{4h}\times{\cal T}$.} 
\label{tab:Dhclassification} 
\end{table*}

For simplicity, we truncate the long-range part of the interaction by taking $r_{\rm max} = \sqrt{2} \times a $.  The interaction is 
then modeled to be 
\bea 
&& V(\pm \hat{x}) = V (\pm \hat{y}) =  V_{\rm nn} \nn \\
&& V(\pm \hat{x} \pm \hat{y} ) = V_{\rm nnn} ^{h}     \nn \\
&& V(\pm \hat {x}  \pm \hat{z}) = V(\pm \hat{y} \pm \hat{z}) = V_{\rm nnn} ^z .  
\eea 
All other terms are taken to vanish. 
With tunability of the lattice geometries, the ratio $V_{\rm nnn} ^z /V_{\rm nn}$ and 
$V_{\rm nnn} ^h /V_{\rm nn} $ is largely tunable between $0$ and $1$. 
The approximate criterion (equation~\eqref{eq:generalcriterion}) for unconventional density waves in this model  becomes 
\be  
3 V_{\rm nn} < 4 V_{\rm nnn} ^h + 8 V_{\rm nnn} ^z . 
\label{eq:pwaveestimate}
\ee 
We then numerically solve self-consistent equations for $p$-wave Rydberg dressed atoms. Both symmetric ( $t_z = t_x = t_y$) and asymmetric ($t_z = 4 t_x = 4t_y$) tunneling cases  are studied.

\begin{figure}[htp] 
\includegraphics[angle=0,width=\linewidth]{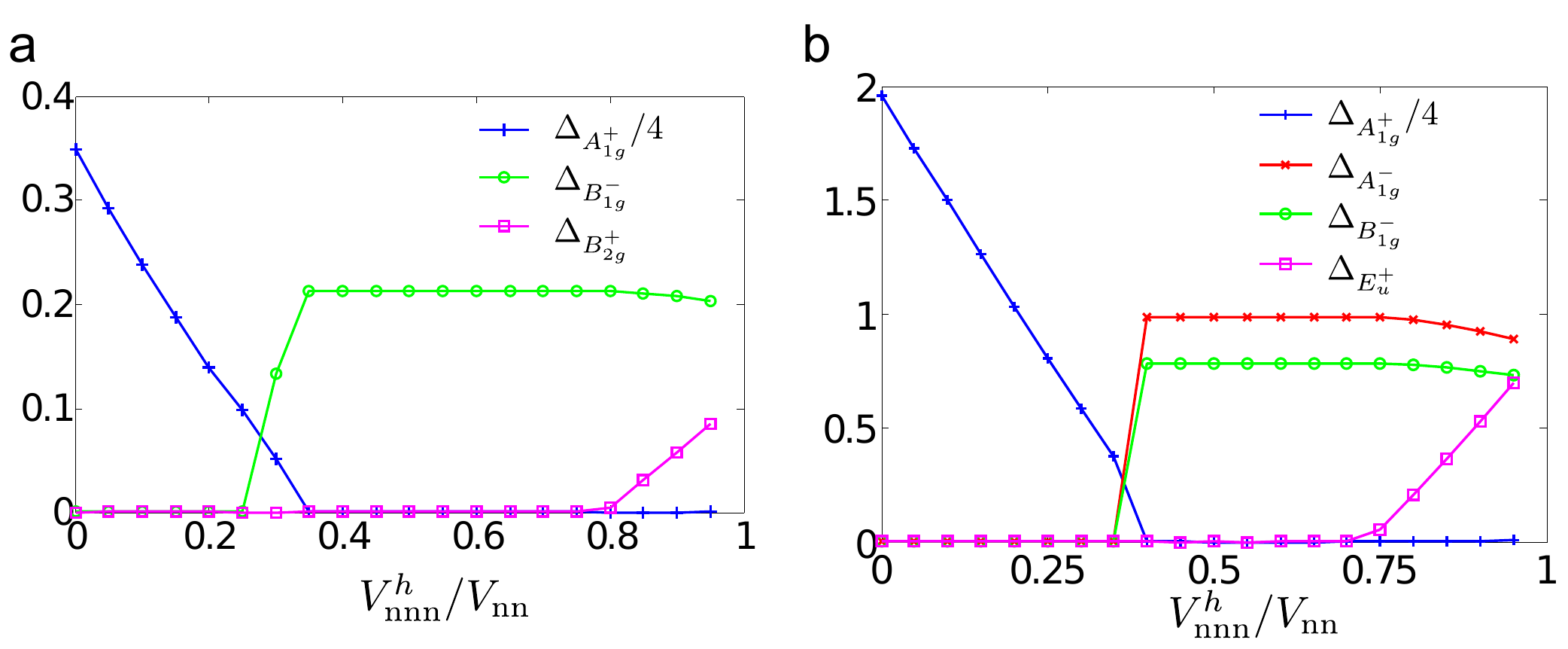}
 \caption{Phase diagram from self-consistent calculations with for $p$-wave Rydberg dressed atoms. In this plot we choose $V_{\rm nn} /t_z = 2$,  $V_{\rm nnn} ^z /V_{\rm nn} = 0.2$. {\bf a} and {\bf b} show the symmetric 
($t_z = t_x= t_y$) and asymmetric ($t_z = 4 t_x = 4 t_z$) tunneling cases, respectively. The $A_{1g} ^+ $ component 
 $\Delta _{A_{1g} ^+ }$ is rescaled by a factor $4$  for illustration purposes.  
 }
\label{fig:phasediagpwave} 
\end{figure}

For the symmetric case (see Supplementary Figure~\ref{fig:phasediagpwave}a), we find $A_{1g}^+$, $A_{1g}^+ + B_{1g}^-$, $B_{1g}^-$ and $B_{1g}^- + B_{2g} ^ +$ density waves. The Fermi surface is fully gapped out for both $A_{1g} ^{+}$ and $A_{1g}^{+} + B_{1g} ^{-}$  states. 
In the  $B_{1g} ^{-}$ state the Fermi surface is only partially gapped out, and we have 
gapless lines (protected by ${\cal T}$ symmetry), from which  the density of states at low energy has linear behavior, i.e., $D(\varepsilon) \propto \varepsilon$. In the $B_{1g}^- + B_{2g} ^+$ state, the Fermi surface for this state is fully gapped out, and 
the complex order $\Delta_{\bf k}$ has a vortex line (with vorticity $2$) elongated along 
the $k_z$ axis. This $B_{1g}^- + B_{2g} ^+$ state is  topologically equivalent to the  discussed $ E_g^- + T_{2g}^+ $ state in the main text. 

For the asymmetric case (see Supplementary Figure~\ref{fig:phasediagpwave}b), we find 
$A_{1g}^+ $, $A_{1g} ^ {-} + B_{1g} ^{-}$, $A_{1g} ^ {-} + B_{1g} ^{-} + E_u^+$ density waves. In the $A_{1g} ^ {-} + B_{1g} ^{-}$ state, ${\cal T}$ is not broken and we have gapless lines. In the  $A_{1g} ^ {-} + B_{1g} ^{-} + E_u^+$ state, 
the density wave order takes a  form, 
\be 
\Delta_{\bf k}  = i \Delta_{A_{1g}^-} (\cos k_x + \cos k_y) 
  + i \Delta_{B_{1g}^{-} } (\cos k_x -\cos k_y) + 
      \Delta_{E_u^{+} } \sin k_x \cos k_y, 
\ee 
which has two straight vortex lines along the $k_z$ direction, given by $(k_x  = \pi/2, k_y = \pm \pi/2)$.  
These vortex lines cross the Fermi surface, and we have four Weyl nodes. Topological numbers for this state are 
$(2,0)$. 

For both cases, the transition from trivial $A_{1g}^+$ state to other density waves roughly occurs at $V_{\rm nnn} ^h/V_{\rm nn}  = 0.35$, with $V_{\rm nnn} ^z /V_{\rm nn}$ fixed at  $0.2$, which agrees with the estimate from equation~\eqref{eq:pwaveestimate}.

\end{document}